Analyzing the Time Evolution of Wave Functions by Decomposing the Hamiltonian into State-Preserving and State-Changing Hamiltonians

Chyi-Lung Lin
Department of Physics, Soochow University,
Taipei 111, Taiwan, R.O.C.

ABSTRACT

We show a new method for analyzing the time evolution of the Schrödinger wave function $\Psi(x, t)$. We propose the decomposition of the Hamiltonian as: $H(t) = \tilde{H}(t) + H_c(t)$, where $\tilde{H}(t)$ is the Hamiltonian such that $\Psi(x, t)$ is its instantaneous eigenfunction, and $H_c(t)$ the Hamiltonian which changes the state $\Psi$. With this decomposition, the action of $H(t)$ on the wave function is simplified and the Schrödinger equation is in a simpler form which can be solved more easily. We illustrate this method by exactly solving the Schrödinger equation for cases of nonspreading wave packets. This method can be applied as well to analyzing the time evolution of general Hamiltonian systems.
.

# 1. Introduction

It is the Hamiltonian that governs the change in time of the wave function $\Psi(x, t)$. This is described in the Schrödinger equation

$$i\hbar \partial_t \Psi(x, t) = H(t) \Psi(x, t) \qquad (1)$$

We discuss the general case that the Hamiltonian may be time dependent; hence the Hamiltonian is denoted by $H(t)$. The action of $H(t)$ on $\Psi(x, t)$ in general is complicate. The simplest case is when $\Psi(x, t)$ is the instantaneous eigenfunction of $H(t)$; then $H(t) \Psi(x, t) = E(t) \Psi(x, t)$, where $E(t)$ is the instantaneous eigenvalue and is a number. In this case, $H(t)$ does not change the state $\Psi$ at the instant t. We can in fact use this concept of instantaneous eigenstate to simplify the action of $H(t)$ on $\Psi(x, t)$, discussed below. The idea is that at each instant t we first determine the Hamiltonian $\widetilde{H}(t)$ such that the wave function $\Psi(x, t)$ is its instantaneous eigenfunction. Then the eigenvalue equation of $\Psi(x, t)$ at each instant is as follows:

$$\widetilde{H}(t) \Psi(x, t) = \widetilde{E}(t) \Psi(x, t) \qquad (2)$$

where we denote the corresponding eigenvalue by $\widetilde{E}(t)$. Eq. (2) tells us the type of the Hamiltonian which does not change the state $\Psi$ at the instant t. We may wonder at first sight how $\widetilde{H}(t)$ can be determined before $\Psi(x, t)$ is solved. In fact, $\widetilde{H}(t)$ can be determined by similarity transformation. We use the concept of time evolution, and we start from an initial wave function $\Psi(x, 0)$. We begin with the relation

$$\Psi(x, t) = U(t, 0) \Psi(x, 0) \qquad (3)$$

where $U(t,0)$ is the time evolution operator. As $U(t,0)$ performs a similarity transformation, formula (2) can be obtained by the similarity transformation from the eigenvalue equation of $\Psi(x, t)$ at t = 0, which is

$$\widetilde{H}(0) \Psi(x, 0) = \widetilde{E}(0) \Psi(x, 0) \qquad (4)$$

For a given $\Psi(x, 0)$, Eq. (4) is obtainable. We then apply the time evolution operator $U(t,0)$ on both sides of (4). In this way we obtain the eigenvalue equation of $\Psi(x, t)$, which is in the form as (2). The corresponding Hamiltonian $\widetilde{H}(t)$ is obtained by the similarity transformation from $\widetilde{H}(0)$. We summarize the results as the following:

$$\widetilde{H}(0)\,\Psi(x,0) = \widetilde{E}(0)\,\Psi(x,0)$$

$\Rightarrow$

$$\widetilde{H}(t)\,\Psi(x,t) = \widetilde{E}(t)\,\Psi(x,t)$$

where

$$\Psi(x,t) = U(t,0)\,\Psi(x,0)$$

$$\widetilde{H}(t) = U(t,0)\,\widetilde{H}(0)\,U^{-1}(t,0) \qquad (5)$$

$$\widetilde{E}(t) = \widetilde{E}(0) \equiv \widetilde{E} \qquad (6)$$

The eigenvalue $\widetilde{E}(t)$ in fact is time-independent, as similarity transformation does not change the eigenvalue. From (5) we can obtain $\widetilde{H}(t)$, and then we may solve $\Psi(x,t)$ from the eigenvalue equation (2); however, we won't solve $\Psi(x,t)$ in this way. Below we show another method for solving $\Psi(x,t)$.

Having determined $\widetilde{H}(t)$ from (5), we let $H_c(t) = H(t) - \widetilde{H}(t)$. The Hamiltonian is then decomposed into

$$H(t) = \widetilde{H}(t) + H_c(t) \qquad (7)$$

Formula (7) means that we divide the Hamiltonian H(t) into two parts. The $\widetilde{H}(t)$ part is the Hamiltonian which does not change the state $\Psi$. The $H_c(t)$ part is the Hamiltonian which actually changes the state $\Psi$. The suffix c is referred to "change". We may call $\widetilde{H}(t)$ the state-preserving Hamiltonian, and $H_c(t)$ the state-changing Hamiltonian. The suffix c may also be referred to "classical", discussed below.

From (7), the action of H(t) on $\Psi(x,t)$ can be simplified as follows:

$$H(t)\,\Psi(x,t) = \left[\,\widetilde{H}(t) + H_c(t)\,\right]\Psi(x,t)$$

$$= \left[\,\widetilde{E}(t) + H_c(t)\,\right]\Psi(x,t) \qquad (8)$$

Schrödinger equation can then be written in a simpler form as below:

$$i\hbar \partial_t \Psi(x,t) = H(t) \Psi(x,t)$$
$$= [\tilde{E}(t) + H_c(t)] \Psi(x,t) \qquad (9)$$

Formula (9) can be solved more easily than (1) or (2). Especially, the time evolution can be understood more clearly via the Hamiltonian $H_c(t)$.

We found that this decomposition method is particularly interesting when it is applied to nonspreading wave packets (NSWPs). There are already known NSWPs constructed by Schrödinger, Senitzky, and also by Berry and Balazs [1-3]. In [4-5], we used these known solutions to determine the corresponding $\tilde{H}(t)$ and $H_c(t)$, and then we use $\tilde{H}(t)$ and $H_c(t)$ to explore the time evolution of these NSWPs. In this paper, we will do in the reverse order, that is, we first determine $\tilde{H}(t)$ and $H_c(t)$, and then we determine $\Psi(x,t)$.

In 1926, Schrödinger constructed the first NSWP with the profile of the ground state of simple harmonic oscillator (SHO) [1]. In 1954 Senitzky generalized Schrödinger's result, constructing NSWPs with the profiles of high energy eigenstates of SHO [2]. Other type of NSWP was found in 1979 by Berry and Balazs [3]. This type of NSWP is in the form of Airy function and occurs in free space and also in a time-varying spatially uniform linear potential. It was found that an Airy packet in free space is not only nonspreading but also self-accelerates. The decomposition of $H(t)$ into $\tilde{H}(t) + H_c(t)$ can offer an explanation to this strange phenomenon. We discuss this phenomenon in Sec. 2.

In Sections 2-4, we apply this decomposition method to solving $\Psi(x,t)$ of NSWPs in free space, in a time-varying spatially uniform linear potential, and in a quadratic potential. In Sec. 5, we make a brief conclusion.

## 2. Airy packets in free space

We first study Airy packets in free space [3]. The Hamiltonian is

$$H = \frac{p^2}{2m} \qquad (10)$$

where $p = -i\hbar \frac{\partial}{\partial x}$. Following Berry and Balazs, the initial wave is the Airy function, i.e., $\Psi(x,0) = \text{Ai}[b\, x]$, where b is an arbitrary constant.

Our goal is to solve $\Psi(x, t)$. The corresponding eigenvalue equation for Ai[b x] is

$$\left(\frac{p^2}{2m} + f_b\, x\right) Ai[b\, x] = 0 \qquad (11)$$

where $f_b \equiv \frac{\hbar^2 b^3}{2m}$. Comparing to (4), we have

$$\tilde{H}(0) = \frac{p^2}{2m} + f_b\, x$$
$$\tilde{E}(0) = 0$$

As $H = \frac{p^2}{2m}$ is time independent, we have $U(t, 0) = \exp\left[\frac{-i}{\hbar} \frac{p^2}{2m} t\right]$. We then obtain

$$U(t, 0)\, x\, U^{-1}(t, 0) = x - \frac{t}{m} p \qquad (12)$$
$$U(t, 0)\, p\, U^{-1}(t, 0) = p \qquad (13)$$

Substituting (12-13) into (5-6), we obtain the eigenvalue equation of $\Psi(x, t)$ as follows:

$$\tilde{H}(t)\, \Psi(x, t) = \tilde{E}(t)\, \Psi(x, t)$$

where

$$\tilde{H}(t) = \left(\frac{p^2}{2m} + f_b\, x - \frac{f_b\, t}{m} p\right) \qquad (14)$$
$$\tilde{E}(t) = 0 \qquad (15)$$

It is interesting to note that although we have not yet solved $\Psi(x, t)$, we know that it is the eigenstate of the Hamiltonian $\tilde{H}(t)$ given in (14) and the corresponding eigenvalue is zero. From (14), we have the decomposition of H as the following:

$$H = \frac{p^2}{2m} = \tilde{H}(t) + H_c(t) \qquad (16)$$

with

$$H_c(t) = \left(\frac{f_b t}{m}p - f_b x\right) \tag{17}$$

From (16-17), and with the result $\widetilde{E}(t) = 0$, Schrödinger equation is now based on the simpler Hamiltonian $H_c(t)$, that is

$$i\hbar \partial_t \Psi(x,t) = H\,\Psi(x,t)$$
$$= H_c(t)\,\Psi(x,t) \tag{18}$$

Eq. (18) is an equation linear in x and p, we can easily solve this equation and obtain

$$\Psi(x,t) = \text{Ai}\left[b\left(x - \frac{f_b t^2}{2m}\right)\right] \exp\left[\frac{i}{\hbar}\left(\frac{f_b t}{m}x - \frac{f_b^2 t^3}{3m}\right)\right] \tag{19}$$

This is the NSWP obtained by Berry and Balazs [1].

The solution of (19) shows that Airy packet self-accelerates in free space with a constant acceleration $a = \frac{f_b}{m}$. According to Ehrenfest's theorem, an NSWP in free space should move with constant speed. However, Airy packet is not square integrable, hence expectation value cannot be defined, and therefore Ehrenfest's theorem cannot be applied. Yet, there still remains the question why an Airy packet self-accelerates in free space. The propagation of a nonspreading Airy packet in free space in fact can be understood from the decomposition of the Hamiltonian shown in (16) and (17). The propagation of a nonspreading quantum packet can be treated as the motion of a classical particle. We note that the classical motion derived from $H_c(t)$ is $\dot{x} = \frac{\partial H_c}{\partial p} = \frac{f_b t}{m}$ and $\dot{p} = -\frac{\partial H_c}{\partial x} = f_b = m\ddot{x}$. This describes a classical motion $x = \frac{f_b t^2}{2m}$ which is just the same as the propagation of the quantum packet. Thus $H_c(t)$ closely connects the propagation of a quantum packet and the corresponding classical motion.

From (19), we note that action of the infinitesimal time evolution operator on $\Psi(x,t)$ is the following:

$$U(t+dt, t)\,\Psi(x,t) = \exp\left[\frac{-i}{\hbar}H(t)\,dt\right]\Psi(x,t)$$
$$= \exp\left[\frac{-i}{\hbar}\left(\widetilde{H}(t) + H_c(t)\right)dt\right]\Psi(x,t)$$
$$= \exp\left[\frac{-i}{\hbar}H_c(t)\,dt\right]\Psi(x,t)$$

$$= \exp\left[\frac{i}{\hbar} f_b \, x \, dt\right] \exp\left[\frac{-i}{\hbar} \frac{f_b t}{m} p \, dt\right] \Psi(x, t) \qquad (20)$$

We see that $U(t + dt, t)$ is effectively a spatial-shift operator, i.e., $\exp\left[\frac{-i}{\hbar} \frac{f_b t}{m} p \, dt\right]$. This shows that $\Psi(x, t)$ in a time interval dt is spatially shifted by an amount of $dx = \frac{f_b t}{m} dt$. Hence an Airy packet moves at a velocity $v = \frac{f_b t}{m}$ and therefore with an acceleration $a = \frac{f_b}{m}$. The reason that Airy packets self-accelerates in free space is because the state-changing Hamiltonian $H_c(t) \neq 0$. For NSWPs, $H_c(t)$ as well as $\widetilde{H}(t)$ can not change the shape of the packet. The effect of $H_c(t)$ is to make a spatial-shift to the packet. The particular form of (17) shows that the spatial-shift operator results an acceleration.

For arbitrary wave packets, $H_c(t)$ does change the shape of packets. That a packet distorts or not in evolution depends on whether $H_c(t) = 0$ or $H_c(t) \neq 0$. The time evolution of a wave function is in fact effectively governed by $H_c(t)$. Free space does not necessary imply $H_c(t) = 0$. Hence a packet in free space does not mean it is free of distortion. In general $H_c(t) \neq 0$; therefore, arbitrary quantum packets distort even in free space. The only exception is Airy packets which do not distort but accelerate.

We may also interpret the phenomenon of self-accelerating in free space by the following viewpoint. From the decomposition formula (16), we may say that the free Hamiltonian $\frac{p^2}{2m}$ offers a part of the Hamiltonian, the state-preserving Hamiltonian $\widetilde{H}(t)$, to maintain the shape of the Airy packet, and then the other part of the Hamiltonian, the state-changing Hamiltonian $H_c(t)$, is then to accelerate the Airy packet. The similar case in classical mechanics is like a bead sliding along a smooth rod [6]. There should be no force in the direction of the rod, because the rod is smooth. But this does not mean the bead will always move at a constant speed. When the rod is beginning to rotate, the bead will be accelerating outward. This is because the bead needs a centripetal force in order to rotate with the rod; the outward force is from the corresponding compensate-force. The mathematical formula of the forces acting on the bead is like the following:

$$0 = \vec{F}_c + \vec{F}_a \qquad (21)$$

where 0 means that there is no net force in the direction of the smooth rod; and $\vec{F}_c$ is the centripetal force needed for the bead to rotate with the rod, and $\vec{F}_a$ is the compensate-force that accelerates the bead outward.

Finally, we note that $H_c(t)$ in formula (17) can be written as follows:

$$H_c(t) = \dot{d}(t)\, p - m\, \ddot{d}(t)\, x \qquad (22)$$

where $d(t) = \frac{f_b\, t^2}{2m}$. We will show in the following sections that this form of $H_c(t)$ is common to NSWPs. More discussions about $H_c(t)$ of NSWPs is referred to [4-5].

## 3. Airy packets in a time-varying spatially uniform linear potential

We next discuss Airy packets in a system with a time dependent Hamiltonian $H(t)$, which is defined as the following:

$$H(t) = \frac{p^2}{2m} - F(t)\, x \qquad (23)$$

where $F(t)$ is an arbitrary function of time [3]. As the commutator $[H(t1), H(t2)] = i\hbar(F(t2) - F(t1))p/m \neq 0$, the time evolution operator should be written as follows:

$$U(t, 0) = \prod_{j=1}^{N} \exp\left[\frac{-i}{\hbar} H(t_j)\, \Delta t\right] \qquad (24)$$

where $t_j = j\, \Delta t$, $\Delta t = t/N$, and N is a large number. The initial wave is $\Psi(x, 0) = \text{Ai}[b\, x]$. The eigenvalue equation of $\Psi(x, 0)$ is then the same as (11). With U(t,0) given in (24), we have

$$U(t, 0)\, x\, U^{-1}(t, 0) = x - \frac{t}{m}\, p + \frac{t\, \alpha(t) - \beta(t)}{m} \qquad (25)$$

$$U(t, 0)\, p\, U^{-1}(t, 0) = p - \alpha(t) \qquad (26)$$

where

$$\alpha(t) = \int_0^t F(t)\, dt \qquad (27)$$

$$\beta(t) = \int_0^t \alpha(t) dt \qquad (28)$$

To express solutions more compactly, we define the following quantity

$$d(t) \equiv \frac{f_b t^2}{2m} + \frac{\beta(t)}{m} \qquad (29)$$

Hence

$$\dot{d}(t) = \frac{f_b t}{m} + \frac{\alpha(t)}{m} \qquad (30)$$

$$\ddot{d}(t) = \frac{f_b}{m} + \frac{F(t)}{m}. \qquad (31)$$

Substituting (25-26) into (5-6), we obtain the eigenvalue equation of $\Psi(x,t)$ and the corresponding $\tilde{H}(t)$ and $\tilde{E}(t)$. We have

$$\tilde{H}(t) \Psi(x,t) = \tilde{E}(t) \Psi(x,t),$$

where

$$\tilde{H}(t) = \frac{p^2}{2m} + f_b x - \dot{d}(t) p - \left(f_b d(t) - \frac{m}{2} \dot{d}(t)^2\right) \qquad (32)$$

$$\tilde{E}(t) = 0 \qquad (33)$$

From (23) and (32), we have

$$H(t) = \frac{p^2}{2m} - F(t) x = \tilde{H}(t) + H_c(t) \qquad (34)$$

$$H_c(t) = \dot{d}(t) p - m \ddot{d}(t) x + \left(f_b d(t) - \frac{m}{2} \dot{d}(t)^2\right) \qquad (35)$$

Using (34-35), and that $\tilde{E}(t) = 0$, Schrödinger equation can be written as follows:

$$i \hbar \partial_t \Psi(x,t) = H(t) \Psi(x,t)$$
$$= H_c(t) \Psi(x,t) \qquad (36)$$

Eq. (36) is again an equation linear in x and p. We can easily solve this equation and obtain

$$\Psi(x,t) = \text{Ai}\,[b\,(x - d(t))]\;\exp\left[\frac{i}{\hbar}\phi(x,t)\right], \qquad (37)$$

$$\phi(x,t) = m\,\dot{d}(t)\,x - \frac{f_b{}^2 t^3}{3m} - \frac{f_b\,t}{m}\int_0^t \alpha(\tau)\,d\tau - \frac{1}{2m}\int_0^t \alpha(\tau)^2 d\tau \qquad (38)$$

This is the NSWP obtained also by Berry and Balazs. We note that the classical motion derived from $H_c(t)$ is $\dot{x} = \frac{\partial H_c}{\partial p} = \dot{d}(t)$ and $\dot{p} = -\frac{\partial H_c}{\partial x} = m\,\ddot{d}(t) = m\,\ddot{x}(t)$. Thus this describes a classical motion $x = d(t)$ which is just the same as the propagation of the quantum Airy packet. Therefore, it is the Hamiltonian $H_c(t)$ connecting the propagation of a quantum packet and the corresponding classical motion.

## 4. Displaced energy eigenstates in SHO

Regarding to SHO, the Hamiltonian is defined as follows

$$H = \frac{p^2}{2m} + \frac{m}{2}\omega^2 x^2 \qquad (39)$$

The Hamiltonian in this case is time independent. We have

$$U(t,0) = \exp\left[\frac{-i}{\hbar} H t\right] \qquad (40)$$

We consider the initial wave as a displaced n-th eigenstate of SHO, that is

$$\Psi(x,0) = \Psi_n(x - d0)\,\exp\left[\frac{i}{\hbar} m\,v0\,x\right] \qquad (41)$$

where $\Psi_n(x)$ is the n-th eigenfunction of SHO with eigenvalue $E_n = (n + \frac{1}{2})\,\hbar$, and d0, v0 are arbitrary constants. We have included the phase factor $\exp\left[\frac{i}{\hbar} m\,v0\,x\right]$ in (41). This is because the solutions of NSWPs in (19) and (38) all have this type of phase factor. The eigenvalue equation of $\Psi(x,0)$ is as the following:

$$\left[\frac{p^2}{2m} + \frac{m}{2}\omega^2(x-d0)^2 - v0\,p\right]\Psi(x,0) = \left(E_n - \frac{m}{2}v0^2\right)\Psi(x,0) \quad (42)$$

Comparing (42) with (4) shows that

$$\tilde{H}(0) = \frac{p^2}{2m} + \frac{m}{2}\omega^2(x-d0)^2 - v0\,p$$

$$\tilde{E}(0) = E_n - \frac{m}{2}v0^2$$

We next determine the eigenvalue equation of $\Psi(x,t)$. From (40), we have

$$U(t,0)\,x\,U^{-1}(t,0) = \cos[\omega t]\,x - \frac{\sin[\omega t]}{m\,\omega}\,p \quad (43)$$

$$U(t,0)\,p\,U^{-1}(t,0) = \cos[\omega t]\,p + m\,\omega\,\sin[\omega t]\,x \quad (44)$$

In order to express solutions in a more compact form, we define

$$d(t) = d0\,\cos[\omega t] + \frac{v0}{\omega}\,\sin[\omega t] \quad (45)$$

Then
$$\dot{d}(t) = v0\,\cos[\omega t] - d0\,\omega\,\sin[\omega t] \quad (46)$$

$$\ddot{d}(t) = -\omega^2\,d0\,\cos[\omega t] - \omega\,v0\,\sin[\omega t] \quad (47)$$

The initial conditions of $d(t)$ and $\dot{d}(t)$ are $d(0) = d0$, and $\dot{d}(0) = v0$. Substituting (43-44) into (5-6), we obtain the eigenvalue equation of $\Psi(x,t)$ and the corresponding $\tilde{H}(t)$ and $\tilde{E}(t)$. We have

$$\tilde{H}(t)\,\Psi(x,t) = \tilde{E}(t)\,\Psi(x,t)$$
with
$$\tilde{H}(t) = \frac{p^2}{2m} + \frac{m}{2}\omega^2 x^2 - \dot{d}(t)\,p + m\,\ddot{d}(t)\,x - \frac{m}{2}\omega^2 d0^2 \quad (48)$$

$$\tilde{E}(t) = E_n - \frac{m}{2}v0^2 \quad (49)$$

From (48), we have the decomposition of the Hamiltonian H as follows

$$H = \frac{p^2}{2m} + \frac{m}{2}\omega^2 x^2 = \tilde{H}(t) + H_c(t) \qquad (50)$$

$$H_c(t) = \dot{d}(t)\, p - m\, \ddot{d}(t)\, x + \frac{m}{2}\omega^2 d0^2 \qquad (51)$$

As a result of (50), Schrödinger equation can be written as the following:

$$i\hbar\, \partial_t \Psi(x,t) = H\, \Psi(x,t)$$
$$= \left(\tilde{E}(t) + H_c(t)\right) \Psi(x,t) \qquad (52)$$

Eq. (52) is again an equation linear in x and p. We easily solve this equation and obtain

$$\Psi(x,t) = \Psi_n(x - d(t))\, \exp\left[\frac{i}{\hbar}\, \phi(x,t)\right], \qquad (53)$$

where

$$\phi(x,t) = m\, \dot{d}(t)\, x - E_n\, t - \int_0^t \left[\frac{m}{2}\dot{d}(\tau)^2 - \frac{m}{2}\omega^2 d(\tau)^2\right] d\tau \qquad (54)$$

This is the NSWP obtained by Senitzky [5]. From (53), we note that the trajectory of the nonspreading quantum packet is $x = d(t)$. We also note that the $d(t)$ in (45) represents the classical motion of a particle in an SHO. However, we may also say that the $d(t)$ represents a classical motion derived from $H_c(t)$, as we have $\dot{x} = \frac{\partial H_c}{\partial p} = \dot{d}(t)$ and $\dot{p} = -\frac{\partial H_c}{\partial x} = m\, \ddot{d}(t) = m\, \ddot{x}(t)$. This describes a classical motion $x = d(t)$, which is the same as the propagation of the quantum packet. Thus $H_c(t)$ is the Hamiltonian connecting quantum mechanics and classical mechanics.

## 5. Conclusion

We analyze the time evolution of wave functions by decomposing the Hamiltonian into a state-preserving Hamiltonian $\tilde{H}(t)$ and a state-changing Hamiltonian $H_c(t)$. Since $\tilde{H}(t)$ does not change the state $\Psi$, Schrödinger equation is essentially based on the Hamiltonian $H_c(t)$. In general, $\widetilde{H}(t)$ contains the operator $\frac{p^2}{2m}$, see Eqs. (14), (32), (48). Therefore $H_c(t)$ is without the operator $\frac{p^2}{2m}$. As a result of that $H_c(t)$ is simpler. We can then solve Schrödinger equation more easily. We illustrate this method by exactly solving the Schrödinger equation for NSWPs constructed by Schrödinger, Senitzky, Berry and Balazs.

The state-preserving Hamiltonian $\widetilde{H}(t)$ can be obtained by similarity transformation from $\widetilde{H}(0)$, which is determined from the initial wave function $\Psi(x, 0)$. The state-changing Hamiltonian $H_c(t)$ is the Hamiltonian which governs the distortion of wave functions. For NSWPs, $H_c(t)$ plays the role as a spatial-shift operator, which then determines the propagation of the packets. We also note that the propagation of nonspreading quantum packets is the same as that of a classical particle whose motion is governed classically by $H_c(t)$. Hence $H_c(t)$ closely connects the propagation of a quantum packet and the corresponding classical motion. The suffix c may also be referred to "classical".

This method can be applied as well to the time evolution of general Hamiltonian systems.

# Reference

1. E. Schrödinger, Naturwissenften **14**, 664 (1926)
2. I. R. Senitzky, "Harmonic Oscillator wave functions", The Phyical Review, **95**, 1115 (1954)
3. M.V. Berry and N.L. Balazs, "Non spreading wave packets", Am. J. Phys. **47**, 264 (1979).
4. Chyi-Lung Lin, Arxiv 1306.1311
5. Chyi-Lung Lin, to appear in Chin. J. Phys. (Taipei)
6. Fowles and Cassiday, "*Analytical Mechanics*", 7$^{th}$ edition, 2005 Brooks/Cole, Cengage Learning, p.200